\documentclass{cernrep} 
\usepackage{texnames}
\usepackage[T1]{fontenc}
\usepackage[bookmarks, colorlinks=true, linktoc=page, pdftex, linkcolor=black, citecolor=black, urlcolor=blue]{hyperref}

\makeatletter
\def\bstctlcite{\@ifnextchar[{\@bstctlcite}{\@bstctlcite[@auxout]}}
\def\@bstctlcite[#1]#2{\@bsphack
  \@for\@citeb:=#2\do{%
    \edef\@citeb{\expandafter\@firstofone\@citeb}%
    \if@filesw\immediate\write\csname #1\endcsname{\string\citation{\@citeb}}\fi}%
  \@esphack}
\makeatother

\usepackage{fancyhdr}
\fancyhfoffset{4 mm}
\fancypagestyle{ARTTITLE}{%
\fancyhf{} 
\lhead{Proceedings of the CERN--Accelerator--School course on
{\it Introduction to Accelerator Physics}}
\lfoot{Available online at \url{https://cas.web.cern.ch/previous-schools}}
\rfoot{\thepage\hspace*{3mm}}
 
}

\begin{document}
\bstctlcite{IEEEexample:BSTcontrol}

\title{Applications of Particle Accelerators}
 
\author {Suzie Sheehy}

\institute{University of Melbourne and University of Oxford}

\begin{abstract}Of the tens of thousands of particle accelerators in operation worldwide, the vast majority are not used for particle physics, but instead for applications. Some applications such as radiotherapy for cancer treatment are well-known, while others are more surprising: food irradiation using electron beams, or the hardening of road tarmac. The uses of particle beams are constantly growing in number including in medicine, industry, security, environment, and cultural heritage preservation. This lecture aims to give a broad sweep of the many uses of particle accelerators, covering technologies ranging in size from a few centimetres for industrial electron linacs through to large synchrotron light sources of hundreds of metres circumference operating as national and international facilities. We finish by discussing some of the challenges facing accelerators used in wider society.
\end{abstract}


\maketitle 
\thispagestyle{ARTTITLE}

\section{Introduction}
\label{sec:intro}

\noindent Particle accelerators have an enormous variety of uses outside of particle physics. It is important that those working in the design, engineering and operation of particle accelerators recognise the role of accelerators and beams in society. Working in the field of accelerator science has the potential for vast impact, not merely restricted to the development of next-generation technologies for particle physics or large scientific installations. This lecture will address a number of main application areas and guide the reader toward additional sources for those we cannot cover in such a condensed overview. A key aim of this lecture is to ensure that those working in the field can address queries from friends, members of the public (and politicians or funding bodies...) about the practical uses of accelerators. \\

\noindent A 2016 report from the USA titled \begin{it} Accelerators for Americas Future\end{it}~\cite{Henning2011} summed up well the enormous variety of uses for particle beams:
\begin{quote}
    A beam of the right particles with the right energy at the right intensity can shrink a tumor, produce cleaner energy, spot suspicious cargo, make a better radial tire, clean up dirty drinking water, map a protein, study a nuclear explosion, design a new drug, make a heat-resistant automotive cable, diagnose a disease, reduce nuclear waste, detect an art forgery, implant ions in a semiconductor, prospect for oil, date an archaeological find, package a Thanksgiving turkey or… discover the secrets of the universe.
\end{quote}
\begin{figure}[h!]
    \centering
        \includegraphics[height=220pt]{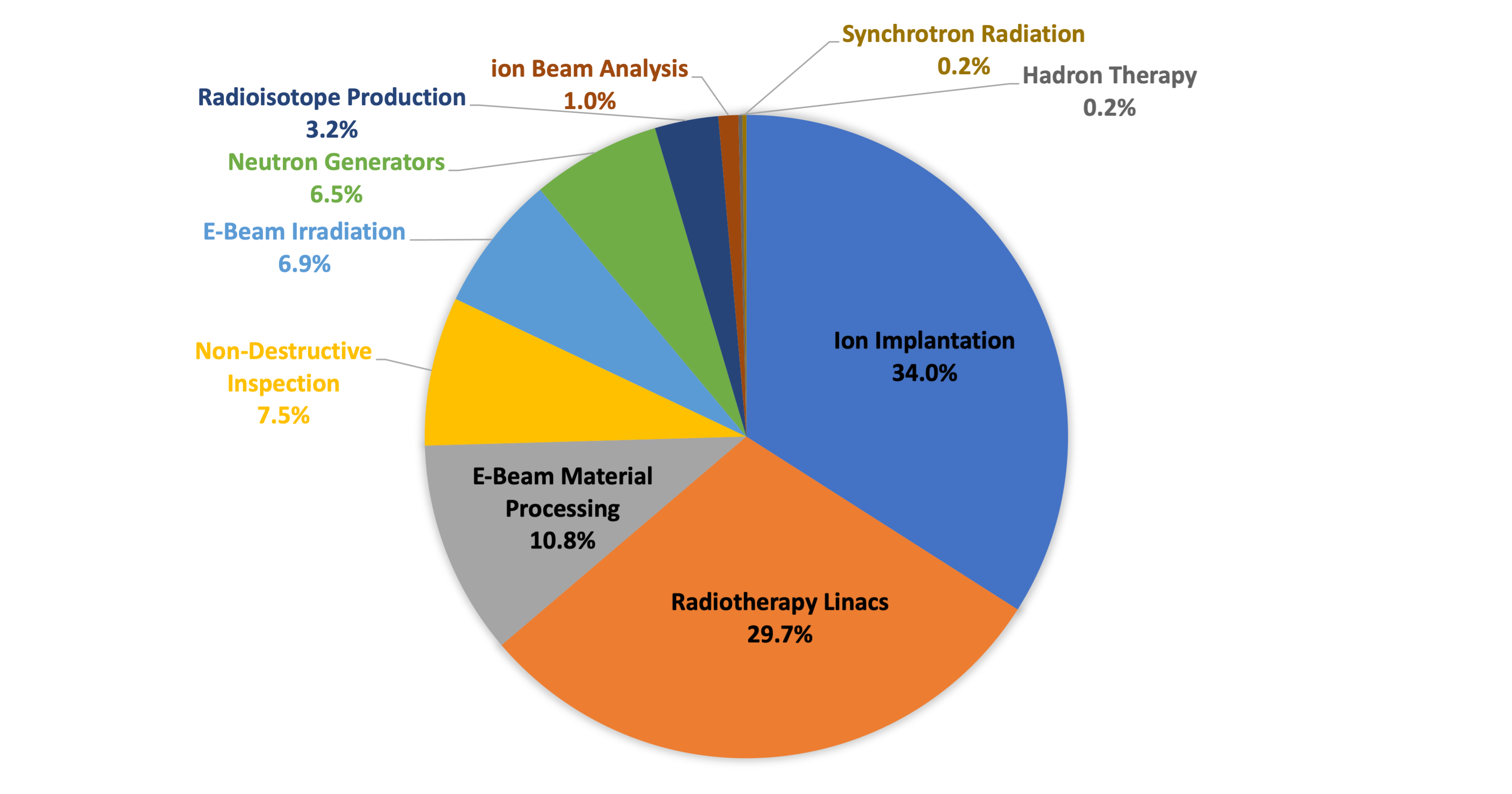}
        \caption{Distribution of accelerators worldwide by common applications in 2019. Data sourced from Ref.\cite{Doyle2018}.}
        \label{fig:1}
\end{figure}

\noindent Let's start with the number of accelerators worldwide: it is estimated that there are over 50,000 particle accelerators in the world\footnote{In this number we do not include cathode ray tube televisions.}~\cite{Doyle2018}. Over half of those accelerators are used for industrial applications, and most of the rest of are used for medical applications~\cite{Hamm2012} as shown in Fig.~\ref{fig:1}. Taken together, accelerators for medical and industrial use comprise an industry that in 2018 was estimated to be worth \$5bn USD per year, growing even during recession. In this overview, we proceed by exploring the most common applications including medical and industrial applications, later moving onto ion beam analysis, synchrotron light sources and neutron sources.

\section{Medical Accelerators}

\subsection{Radiotherapy}
\begin{figure}[h!]
    \centering
        \includegraphics[height=220pt]{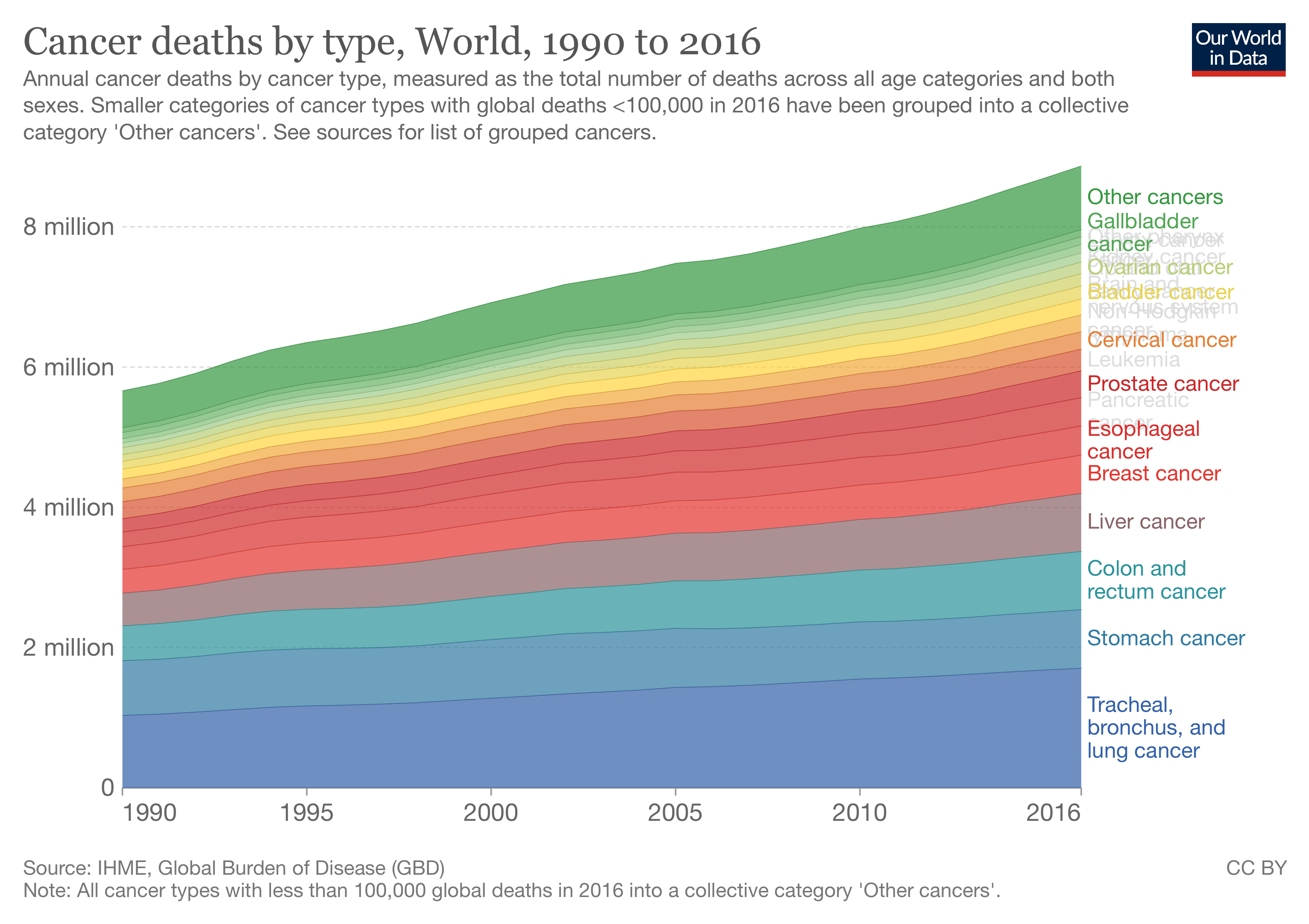}
        \caption{Figure showing overall increase in cancer deaths over time by cancer type, from OurWorldinData under CC-BY license.}
        \label{fig:2}
\end{figure}

Cancer is the second largest cause of death worldwide after cardiovascular diseases. Thankfully, we are getting better at treating cancer, so the age-standardised cancer death rate is decreasing and on average people are living longer and healthier lives than ever before. Despite this, the population is increasing and so overall the number of global cancer deaths is increasing, shown in Fig.~\ref{fig:2}.

\noindent Radiotherapy is used in around half of all curative cancer cases where it is available. Radiotherapy directs ionising radiation of X-rays or electrons toward cancerous regions (tumours) which causes DNA damage in the cancer cells and leads to cell death. To spare healthy tissues, beams of radiation must be shaped to conform to the treatment volume, and the aim is to have a high therapeutic ratio: that is, a favourable ratio between probability of tumour control and toxicity based on normal tissue damage. Since it's inception in the 1950s radiotherapy has been revolutionised through advances in medical imaging, computing, and technological advances leading to new treatment paradigms including Intensity Modulated Radiation Therapy (IMRT) and Volumetric Modulated Arc Therapy (VMAT). It is used routinely in High Income Countries (the same cannot always be said for Low and Middle Income countries, we will address this in later in `challenges'). 

\noindent Today there are around 12,500 radiotherapy linacs in the world, typically S-band (3\,GHz) linear accelerators of either standing wave or travelling wave type. The accelerator itself is compact: usually between 30 to 100cm long. The beam is first generated in a thermionic cathode, then captured and accelerated by the first few cells of the linac and accelerated to an energy typically between 6 and 25 MeV. After this, a magnetic achromat guides the electron beam toward a metal target, where the high energy electrons produce bremsstrahlung X-rays. Electrons can be used directly in certain cases, usually enabled by a retractable target. The most typical treatments involve X-rays, which are collimated and directed toward the patient. 

\begin{figure}[h!]
    \centering
        \includegraphics[height=140pt]{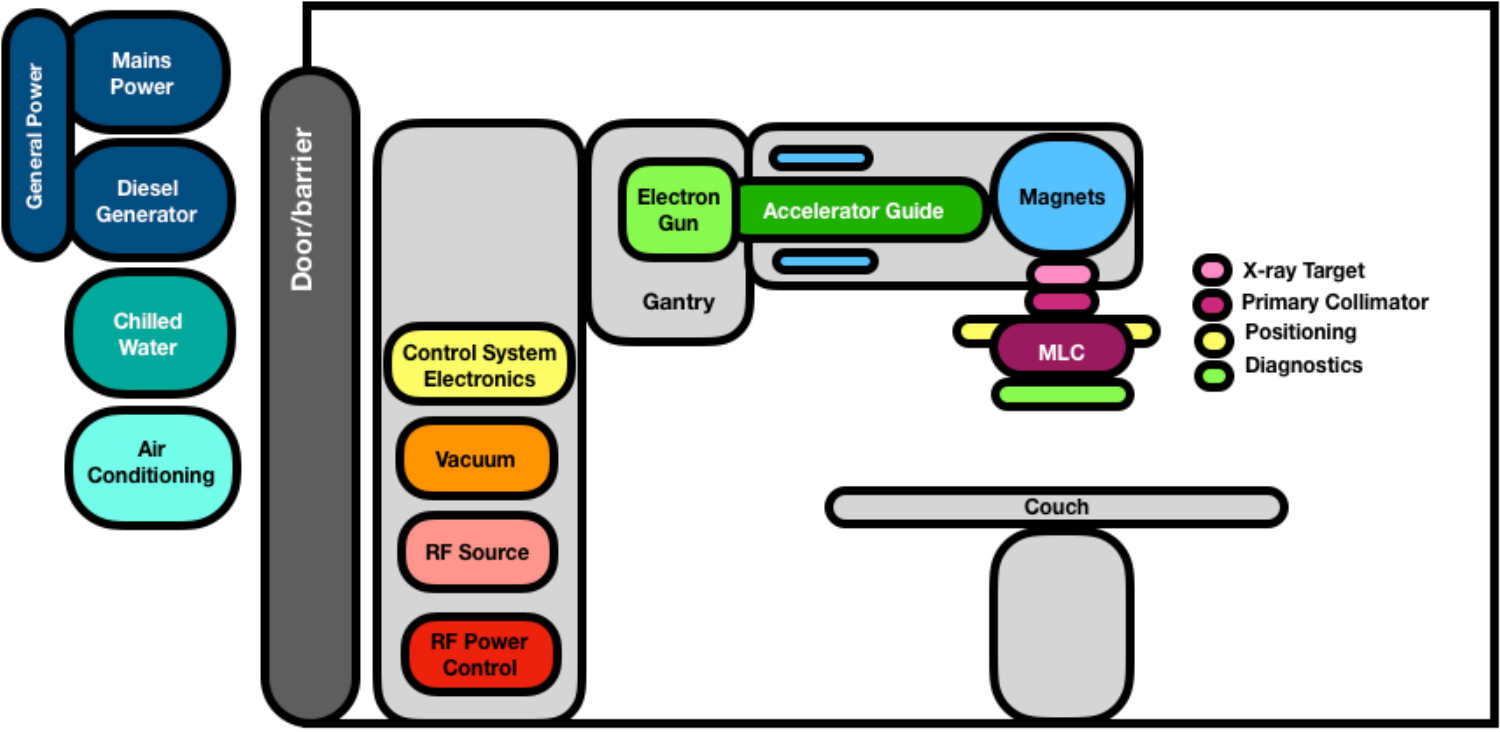}
        \caption{Simplified block diagram of radiotherapy LINAC indicating different subsystems.}
        \label{fig:3}
\end{figure}

\noindent The linac is typically mounted on a gantry, a mechanical structure which rotates 360 degrees around a patient to deliver beams from any angle. This multi-angle delivery technique allows better tumour conformity, better sparing of sensitive organs and increased therapeutic ratio. It can be used either in a 'step-and-shoot' mode (in IMRT) or in a continuous delivery mode (usually VMAT) where the machine is moving while delivering radiation. A basic layout of a medical LINAC integrated into a clinical system is shown in Fig.~\ref{fig:3}. For further reading, a thorough overview of medical linacs can be found in Karzmark et al.,\cite{Karzmark2018, Karzmark1993}, while more up-to-date overviews are given in books by Hanna on RF linear accelerators~\cite{Hanna2012} and a text by Green and William~\cite{Greene2017}.

\subsection{Particle therapy}
In 1946 at the same time that X-ray based radiation therapy was still in its infancy, particle accelerators for protons and ions first reached energy levels of a few hundred MeV. Robert Wilson (later director of Fermilab) was the first person to realise that because these particles would reach deep inside the human body, this might make heavy charged particles of interest in cancer treatment~\cite{Wilson1946}. Many decades of research, development and clinical trials ensued, until today more than 100 facilities worldwide treat patients with protons and in some cases heavier ions, predominantly Carbon~\cite{PTCOG-web}.

\begin{figure}[h!]
    \centering
        \includegraphics[height=180pt]{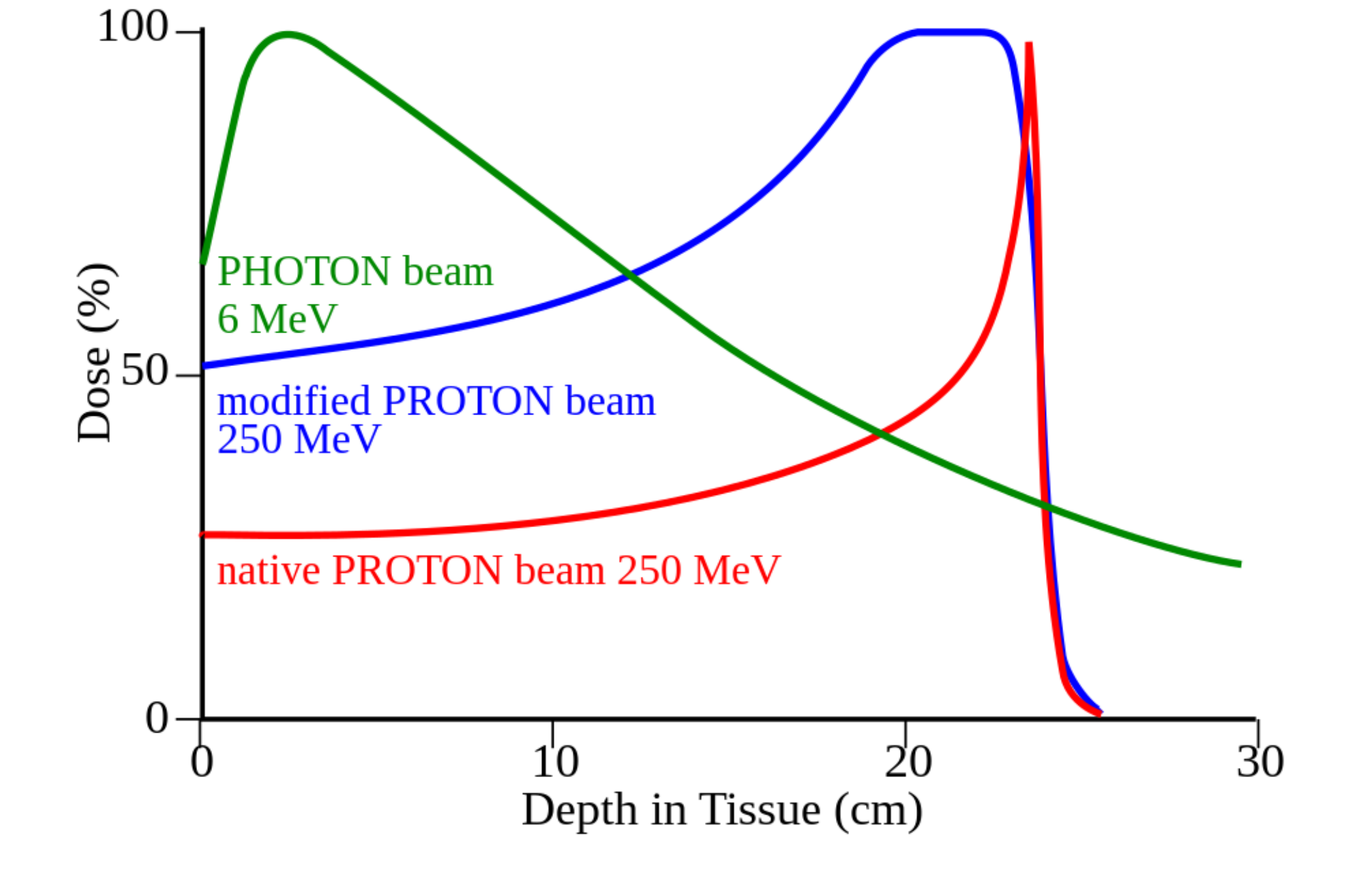}
        \caption{Comparison of depth-dose curve produced by X-rays and protons, showing the characteristic `Bragg Peak'. In practice, a Spread Out Bragg Peak is used, shown as the modified curve.}
        \label{fig:4}
\end{figure}

\noindent Proton therapy utilises a beam of protons generated by a particle accelerator to target tumour cells precisely while minimising damage to surrounding healthy tissue. Unlike conventional radiotherapy methods, where MV energy range X-rays pass through the patient's body and can cause damage along their path, proton beams deposit most of their energy at the precise location where they come to a stop. This difference of radiation interaction in matter is governed by the Bethe-Bloch equation and results in the `Bragg peak' shown in Fig~\ref{fig:4}. This targeted approach has the potential to significantly reduce side effects and improve treatment outcomes.

\noindent The required maximum beam energies to reach sufficient depth in tissue to treat any site (generally taken as around 25cm depth) corresponds to approximately 250~MeV protons or 430~MeV/u carbon ions. Helium, Oxygen and other particles species are also under active investigation. The high rigidity of the required particles makes particle therapy accelerators and all related ancillary systems including beam delivery systems and gantries much larger than for X-ray radiotherapy. Commercial accelerator systems exist for both protons and heavy ions from a number of vendors. Hospital-based systems can be procured which range from single room single-gantry systems driven by a compact proton cyclotron, through to large facilities with multiple gantries and capability for multiple ion species, typically driven by a synchrotron. For further information, we refer to a review of the history of hadron therapy accelerators~\cite{Degiovanni2015}, and for the accelerator technology side, further information can be found in a 2016 review of hadron therapy accelerators~\cite{Owen2016}.

\noindent The particle therapy accelerator field continues to innovate, with new accelerator technologies for accelerators, gantries and beam delivery systems constantly being developed in order to improve, scale down, or speed up particle therapy. Development of commercial linear accelerators in this application space is ongoing. Many design studies have also been carried out on the use of fixed field accelerators (FFAs) for particle therapy, with a number of prototypes built. The use of FFA optics is also under development for improved beam delivery systems to alleviate limitations in the time taken to switch energy layers, which can dominate total treatment time~\cite{Yap2021}. Laser-plasma accelerators are also being actively explored as future options for particle therapy, with experimental facilities actively exploring the radiobiology of laser-plasma driven ions.

\noindent For completeness, it should also be noted that a large body of recent work has explored the use of Very High Energy Electrons in the 100 - 250 MeV range for therapy. While not yet in clinical use, this option may provide an alternative to both radiotherapy and proton/ion therapy in future~\cite{Ronga2021, Whitmore2021} as it provides deeply penetrating beams with a lower entrance dose when used in focused mode, and compared to heavier charged particles electrons appear to be relatively insensitive to inhomogeneities in the body which affect accuracy of dose deposition particularly when a patient has anatomical changes between fractions, or during treatments where organ motion is a factor. 

\subsection{Radioisotopes and other medical accelerators}
Accelerators (compact cyclotrons or occasionally linacs) are essential tools in the production of radio-isotopes for medical imaging (and other applications). These are typically generated in two energy ranges: either low-energy 7-11\,MeV protons for short-lived isotopes for imaging, or 70-100\,MeV or higher beam energies for longer lived isotopes. Isotopes are typically generated using proton, deuteron, 3He and 4He beams. 

\noindent One of the most common procedures for medical imaging employs the use of the radioisotope Fluorine-18 in PET scans (Positron emission Tomography). Fluorine-18 has a half life of ~110 min. Once produced, the Fluorine-18 is chemically attached to a sugary liquid, forming Fluorodeoxyglucose or FDG. When this is delivered to a patient, it carries the F-18 to areas of high metabolic activity. F-18 is a positron emitter: when the positron comes into contact with an electron locally inside the body it annihilates, producing two photons emitted back-to-back (thanks to energy and momentum conservation). These two photons are then detected using a ring of detectors around the patient. In this way, an image can be accumulated, which reveals not the physical structure, but the metabolic activity inside the patient. 

\noindent Around 90\% of PET (Positron Emission Tomography) scans are in clinical oncology. A range of other isotopes are used for PET, as well as for SPECT (Single Photon Emission Computed Tomography) and Brachytherapy for direct treatment. Some of these key uses of radioisotopes in the medical field are summarised in Table~\ref{tab:radioisotopes}.

\begin{table}[h]
\begin{center}
\caption{Common medical radio-pharmaceuticals, reproduced from ANSTO~\cite{ANSTO-isotopes}}
\label{tab:radioisotopes}
\begin{tabular}{llp{10cm}}
\hline\hline
\textbf{Isotope} & \textbf{Half Life} & \textbf{Use}\\
\hline
Carbon-11   & 20.33 minutes            & Used in Positron Emission Tomography (PET) scans to study brain physiology and pathology, to detect the location of epileptic foci, and in dementia, psychiatry, and neuropharmacology studies. Also used to detect heart problems and diagnose certain types of cancer. \\
Nitrogen-13     & 9.97 minutes             & Used in PET scans as a blood flow tracer and in cardiac studies. \\
Oxygen-15 & 2.04 minutes & Used in PET scans to label oxygen, carbon dioxide and water in order to measure blood flow,  blood volume, and oxygen consumption. \\
Fluorine-18 & 1.83 hours & The most widely-used PET radioisotope. Used in a variety of research and diagnostic applications, including the labelling of glucose (as fluorodeoxyglucose) to detect brain tumours via increased glucose metabolism.\\
Copper-64 & 12.7 hours & Used to study genetic disease affecting copper metabolism, in PET scans, and also has potential therapeutic uses. \\
Gallium-67 & 78.28 hours & Used in imaging to detect tumours and infections.\\
Iodine-123 & 13.22 hours & Used in imaging to monitor thyroid function and detect adrenal dysfunction.\\
Thallium-201 & 73.01 hours & Used in imaging to detect the location of the damaged heart muscle.\\
\hline\hline
\end{tabular}
\end{center}
\end{table}

\noindent Accelerators are also used for a type of radiation therapy known as Boron Neutron Capture Therapy (BNCT). This is a combined therapy, where a patient is first injected with a non-radioactive isotope Boron-10 which is 'tumour seeking' and has a high neutron cross section. A high current accelerator is used to generate a beam of low energy neutrons, which then interact with the Boron and a nuclear reaction creates short-range alpha particles, which only reach a few micrometres from their point of origin, thus providing a very localised dose to the tumour. While it is not yet widespread, a number of commercial systems are now available for this therapy. Further information on the radiobiology of BNCT can be found in \cite{Coderre1999}.

\noindent Finally, there exist many other medical uses of radiation that could benefit from using accelerators rather than radioactive sources. For example in some cases, blood used for transfusions is irradiated in order to prevent Transfusion-associated graft-versus host disease (TA-GVHD) associated with lymphocytes, however the appropriate machine is often not available in Low and Middle-Income Countries. It could be more effective to use a radiotherapy LINAC for this purpose, depending on the patient load of the radiotherapy department~\cite{Shastry2013}.

\section{Industrial uses of accelerators}
Direct beams of electrons and ions are used along with secondary beams: neutrons or photons, in a dazzling array of applications in a number of industry sectors: from material processing, to non-destructive testing and analysis. These accelerator based techniques are typically used to speed up or enable processes, manufacturing techniques or treatments that are otherwise impossible.

\subsection{Ion implantation for semiconductors}
\begin{figure}
    \centering
    \includegraphics[height=180pt]{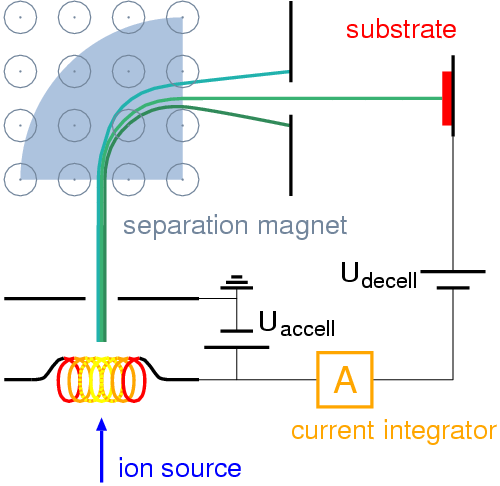}
    \caption{Ion implanter schematic (Image: Creative Commons)}
    \label{fig:5}
\end{figure}

The most common use in industry of ions is in the precise implantation of dopants in semiconductors, which results in the precise modification of electrical, optical and other properties: this can enhance device performance and create new functionalities~\cite{Felch2013}. The use of semiconductor based devices is growing, they are found in computers, smartphones and devices like TVs, but nowadays there are more semiconductor devices in your car than your computer. Accelerators provide ions typically in the hundreds of eV to MeV energy range, and a separation magnet is often used to select ion species at a specific energy. As well as a large range of energies, ion implantation also employs a large range of beam currents. A rough schematic of an ion implanter system is shown in Fig.\ref{fig:5}. 

\begin{figure}[!t]
    \centering
    \includegraphics[height=200pt]{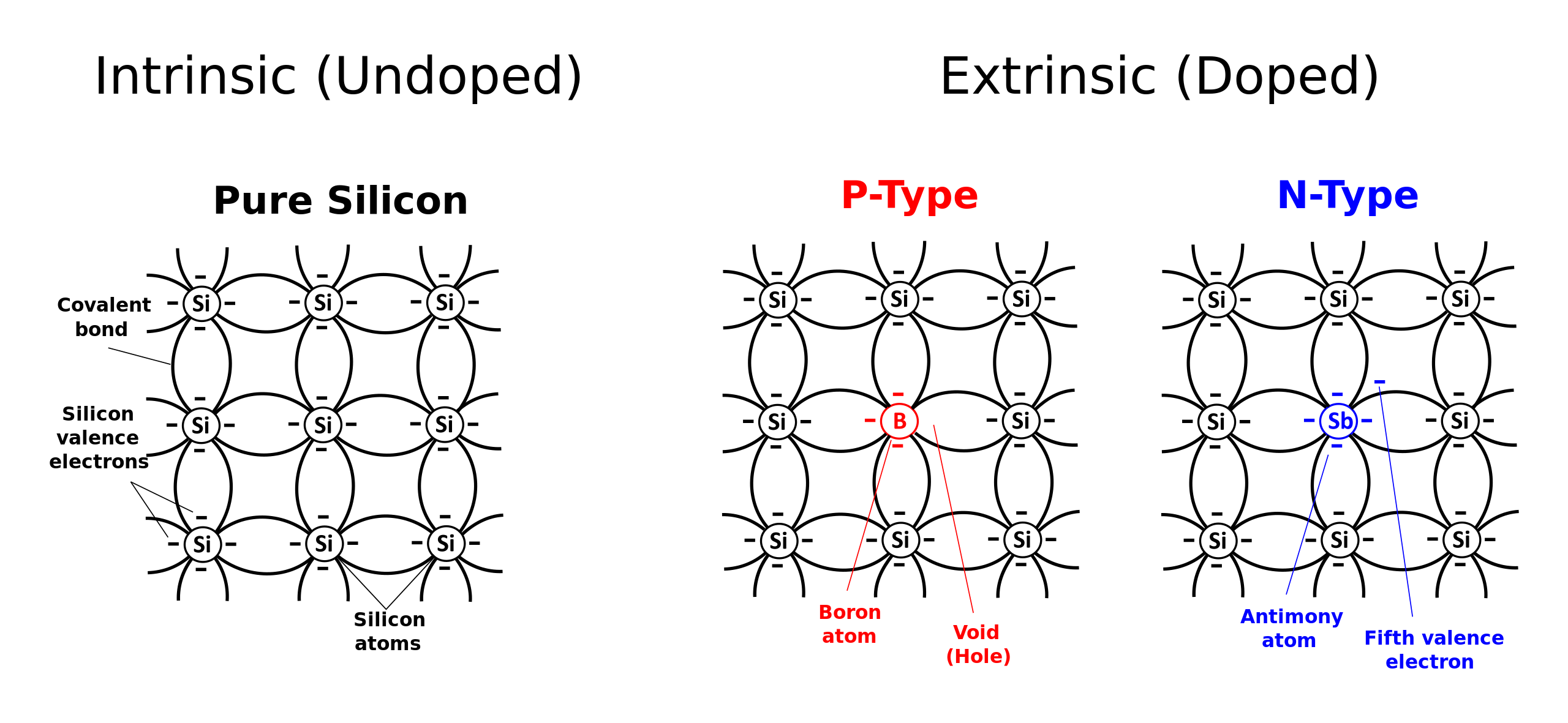}
    \caption{Diagram of P and N type silicon doping using Boron and Antimony. Image: CC BY-SA 4.0 by VectorVoyager}
    \label{fig:6}
\end{figure}

\begin{figure}[h]
    \centering
    \includegraphics[height=200pt]{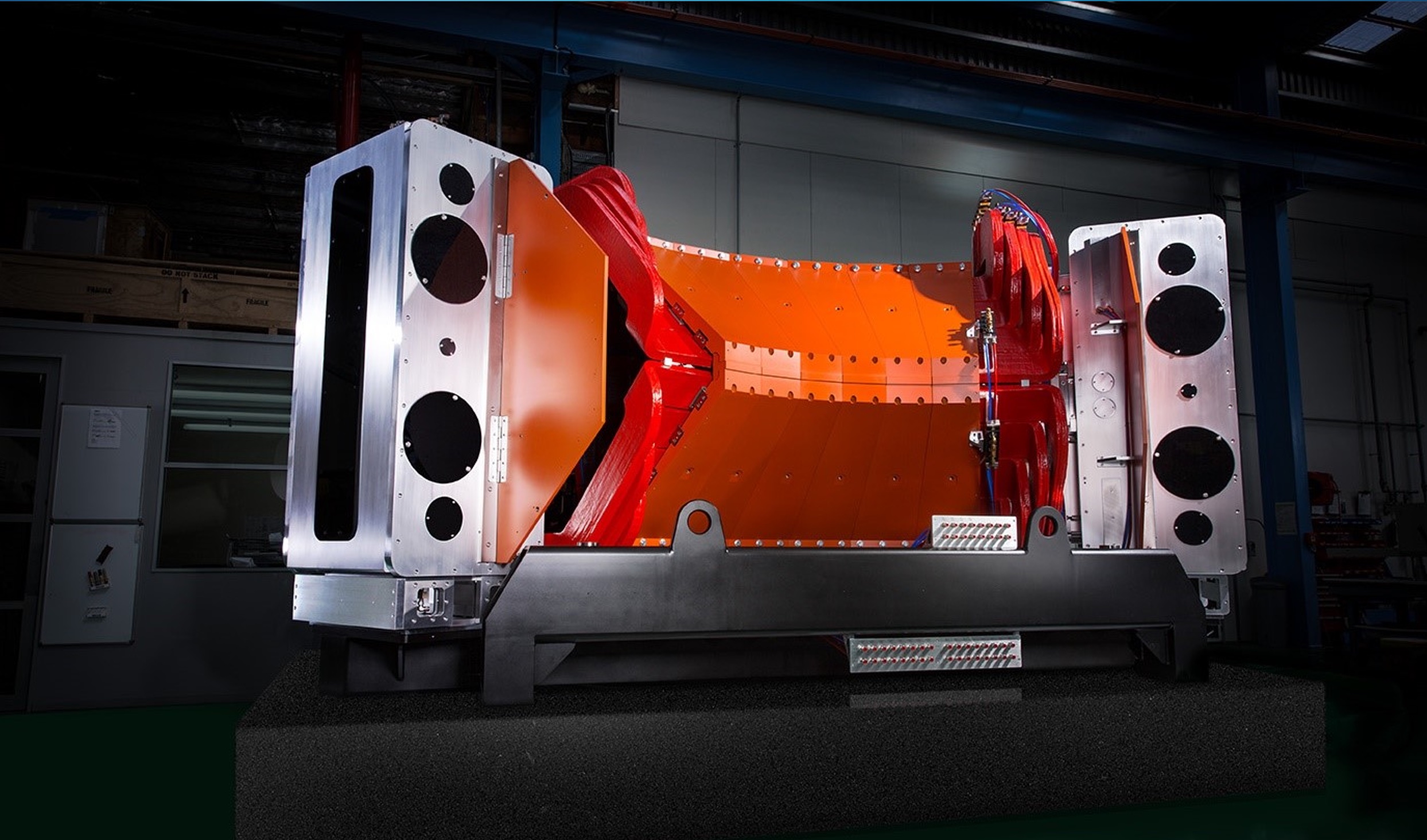}
    \caption{A magnet manufactured for semiconductor industry, Image: Buckley Systems.}
    \label{fig:7}
\end{figure}

The ion implantation method allows precise control over dopants down to the single ion level, as illustrated in Fig.~\ref{fig:6}, which is not possible with other methods such as diffusion. The most common dopants used in the semiconductor industry are Boron, Antimony and Phosphorus. This method is highly selective so can be used to dope specific sub-regions of a material. The success of ion implantation has been one of the technologies underpinning the enormous growth in computing power in recent decades, but also extends beyond this: ion implantation has been used to produce modern transistors, LEDs and photovoltaic cells, among other devices. 
There are around 12,000 ion implantation devices in operation, with a 2009 paper~\cite{Felch2013} estimating the sale of ion implantation devices was even then a 1.5 billion USD market per year, and growing. Over time, the trend in ion implantation systems is for them to become more precise, more sophisticated and more reliable. This involves, for instance, the development of precision electromagnets specific to this application, as shown in Fig.~\ref{fig:7}. Further information about the use of accelerators in the semiconductor industry can be found in \cite{Larson2012, Current2012}.

\subsection{Electron beam irradiation}
Electron irradiation can modify both chemical and material properties of a target substance. Applications include ink curing, adhesives, sterilisation of medical products, disinfection and food preservation. New applications are also emerging including treatment of waste water and gases from industrial processes. Many of the materials routinely irradiated are polymers: when polymers are cross-linked, they can become stable against heat, have increased tensile strength, and develop resistance to cracking or heat shrinking properties. Polymer cross-linking applications include: wire cable tubing, ink curing, shrink film, tyres and other areas~\cite{Sabharwal2013}. Electron beams can even be used to cure or harden the tarmac surface of roads.

\noindent In a typical electron irradiator facility, a high current but modest energy (MeV range) electron accelerator is used to create a fan-shaped beam of electrons, through which a conveyor belt moves the product to be irradiated. The beam may either be scanned rapidly, or non-scanned. The throughput and irradiation rate relies on high beam current, up to multiple mA with a total beam power reaching hundreds of kilowatts for commercial industrial systems.\cite{Reed2012} In the US alone, potential markets for industrial electron beams total 50 billion USD per year. Electron beam irradiation is considered to be an environmentally friendly option compared to 'wet' or chemical processing methods~\cite{Sabharwal2013}.

\noindent For an example of a relatively common object that is frequently irradiated, many people need to look no further than their own hand on neck as they are wearing gemstones which have been irradiated. Kunzite, tourmaline, topaz, quartz, aquamarine, cultured pearls and even diamonds are routinely irradiated. Electron beam irradiation typically lower than 9 MeV can lead to color enhancement and improved clarity, by knocking off electrons in the material thus generating new color centres which result in a deepening of the colour. Doses required are high: in the kGy to hundreds of kGy region~\cite{Idris2012}.

\noindent Another example of the ubiquity of irradiated materials can be found in the analysis of a car. A breakdown of objects in a typical vehicle which can be modified or enhanced using particle beams reveals a staggering array of material modification applications of mostly electron beams: ``components such as tyres, foam, ball bearings, gears, camshafts and tie-rod ends are produced using either electron beam thermal processing or irradiation. Modern ion implantation systems (see later) make the advanced electronic systems in cars possible"~\cite{AccSoc}.

\subsubsection{Sterilisation techniques}
Electron beams can be used as a sterilisation technique, and this is used extensively in industries such as medical equipment sterilisation, food processing and pharmaceuticals. Manufacturers of medical disposables have to ensure that syringes, bandages, surgical tools and other gear do not carry pathogens, without altering the material itself. Between 40-50\% of all disposable medical products manufactured in North America are currently radiation sterilised, primarily at 60Co irradiation facilities\cite{Doyle2018}. When an electron beam is used instead, it is referred to as `e-beam sterilisation'. 

\noindent Compared to gamma irradiation from a sealed radioactive source, e-beam methods are typically faster, of the order of a few seconds rather than hours. However, this comes at a cost of electron beams typically having limited penetration depth particularly through dense material. This makes e-beam processing best for simple, low density products. Many medical supplies fall into this category.

\subsubsection{Wastewater irradiation}
It has been claimed that ``More people die from contaminated water in this world than from any disease"~\cite{SymmetryElectrons}. At the same time, increasing consciousness of the wider environmental impacts of chemical effluents from industries such as large-scale dyeing, means that electron beams have found a growing application in the processing of wastewater. Further uses of this technique to remove organic compounds and disinfect or reclaim wastewater can be applied in municipal wastewater, contaminated underground water, pharmaceutical and petrochemical industries.

\noindent When irradiated, short-lived reactive radicals are used to oxidise or reduce pollutants into a liquid waste form without introducing harmful byproducts. The resulting liquids can often be used as fertilizer. These clean and sustainable methods address the growing concerns surrounding water pollution and resource scarcity. An early example of this was in Korea at the Daegu Dyeing Complex (DYETEC), where today a 10,000m3/day plants is in successful operation and further facilities are now found in Sri Lanka and Brazil~\cite{Han2005}.

\noindent Accelerators for wastewater irradiation are similar in parameters to the factory-based electron beam irradiators, requiring a 1 MeV, high current scanning system. Reduction of certain chemicals can depend strongly on the dose. A dose of around 1 kGy is used to remove color, odor, disinfection and removal of some organic pollutants. 

\noindent A challenge for these accelerators is reliability, size and efficiency, and although progress has been slow there are a number of operational systems in parts of Europe, Asia and the Middle East. In recent years, new projects have been launched to design accelerators specifically for environmental remediation, including both normal conducting linac and superconducting RF options.~\cite{JLAB}

\subsubsection{Food irradiation}
\noindent Food irradiation often strikes people as an unexpected application of electron beams, yet in a world with increasing focus on food security and global supply chains, it has been steadily growing in use. Irradiating food does not affect its nutritional quality, and does not leave residual radiation: rather, it is a technique which generally aims to prevent pathogen contamination of foods. Techniques using either X-rays or electrons to treat food are referred to as `nonthermal' food processing or in some cases `cold pasteurisation' or `electronic pasteurisation'. Again, because electron beam irradiation effectively eliminates pathogens, this can prolong the shelf life of food products while preserving their nutritional value. 

\noindent For example, in Australia in 2021 the Queensland Government was approved to irradiate all types of fresh fruit and vegetables for phytosanitary reasons, to prevent the spread of fruit fly. It is only proposed to irradiate up to a few percent of all fresh fruit and vegetables, but all types have been approved. 

\begin{figure}[h]
    \centering
    \includegraphics[height=50pt]{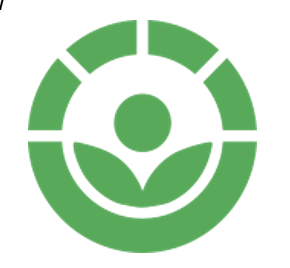}
    \caption{The international 'Radura' symbol means that a food product has been irradiated. In most countries, a statement must also appear on the packaging.}
    \label{fig:8}
\end{figure}

\noindent In the European Union, only dried herbs and spices are currently approved for irradiation and the words `irradiated' or `treated with ionising radiation' must appear on the label packaging. There is an international Radura symbol, shown in Fig.\ref{fig:8} which also indicates irradiated food, however this author has never seen it on a food label. In the USA and Canada a wider range of foods are approved including beef for microbial decontamination, poultry to control food-borne pathogens, fresh foods to control microorganisms and eggs for control of salmonella~\cite{FSANZ}.

\subsection{Non-destructive testing}
Non-destructive testing frequently involves electron linear accelerators up to a few MeV in beam energy. These electrons are used to generate X-rays for CT inspection of high-density items or objects which are large and difficult to inspect in other ways. Typical objects include cast or forged items, turbine blades, aerospace or aircraft components including engines or defence items such as missiles.\cite{Oriatron}.

\noindent Non-destructive testing also finds increasing use in civil engineering applications, allowing engineers to identify internal defects in structures such as pipelines, buildings, bridges, garages and other objects~\cite{Owen1998}. These techniques are becoming increasingly important in an era of 3D printing where internal defects can arise and be harder to characterise than traditional manufacturing techniques. Further reading can be found in~\cite{Reed2012}.

\section{Ion Beam Analysis techniques}
A suite of techniques come under the description of ion beam analysis, typically using light ions such as protons or Helium in the few MeV range to analyse the elemental composition in the first few layers of the surface of a material. IBA techniques can also be used to modify materials, including the implantation of ions in cutting-edge silicon and other devices such as qubits for quantum computing.

\noindent Common techniques include Rutherford Backscatter (RBS), where the elastically scattered ions are detected and used to determine elements at shallow depth, which is most sensitive to heavy elements in a light material, and Particle Induced X-ray Emission (PIXE) in which atomic interactions generate bursts of characteristic X-rays (or gamma rays in the case of PIGE), creating a ‘fingerprint’ for elemental analysis. PIXE is a commonly used technique among cultural heritage and art conservators in determining authenticity.  

\begin{figure}[h]
    \centering
    \includegraphics[height=200pt]{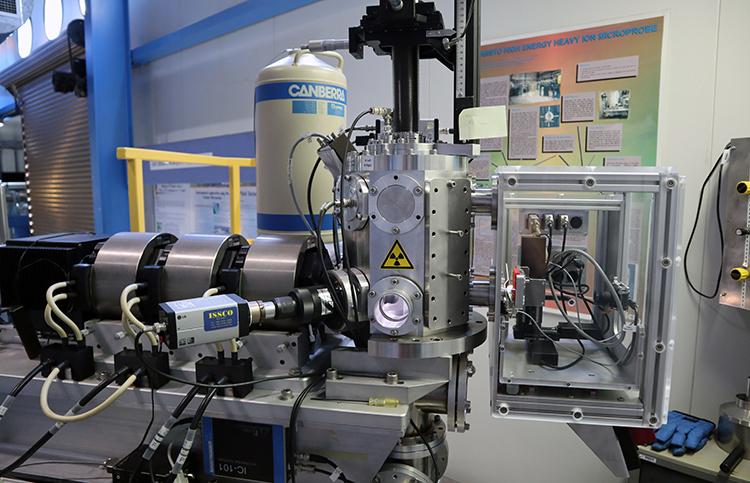}
    \caption{The high energy heavy ion microprobe at ANSTO, Sydney, Australia (Image: ANSTO).}
    \label{fig:9}
\end{figure}

\noindent There are almost endless applications of IBA techniques. According to the IAEA Accelerator Knowledge Portal\cite{IAEA}, example of application areas include:
\begin{itemize}
    \item Nanotechnology, such as the creation of nanofabricated structures.
    \item Semiconductors and electronic devices, for instance by ion implantation, including quantum devices.
    \item DNA modification, for instance mutagenic breeding of plants. 
    \item Determining the origin of pollutants such as fine aerosols in air or sediment particles transported by water.
    \item Imaging individual biological cells.
    \item Determining the trace elements in tissues and understanding mechanisms of disease.
    \item In art and archaeology for determining the composition of inks, paints and glazes on ceramics and glasses to confirm their origin.
\end{itemize}

\subsection{Case Study: Air pollution monitoring}
\noindent One case study is in the contribution of IBA techniques using particle accelerators to environmental studies and solutions. According to David Cohen at ANSTO, Australia, an expert in these techniques: “Using ion beam analysis methods, it is possible to not only determine the number of minute quantities of pollutants in the air but also to identify their sources". In air pollution monitoring, accelerator-driven techniques such as accelerator mass spectrometry enable researchers to analyse aerosol particles in great detail. This information aids in understanding the sources and composition of pollutants, guiding efforts towards effective mitigation strategies. In a study in Australia it was found that “up to half of the total sulfate air pollution in the greater Sydney region can be attributed to emissions from NSW’s eight coal-fired power stations”~\cite{Cohen2004}.

\subsection{Case Study 2: Food Provenance}
\noindent A further emerging application is in the domain of food provenance. Ion beam analysis can be used to determine the origin of food both from plants and animals including fish. The place where the animal or plant grew creates a unique environmental fingerprint after taking in elements and their isotopes from local sources of water and nutrients. Ion beam analysis can identify these isotopes and elements including their relative ratios, allowing scientists to link a food product to its origin. This is of particular importance in verifying the authenticity of food origin~\cite{ANSTO-food}.

\subsection{Radiocarbon and Radiometric dating}
Accelerator Mass Spectrometry (AMS) is a highly sensitive technique used to measure long-lived radionuclides with high precision. The process begins with ionising the sample and selecting specific ions, which are then accelerated and passed through a series of magnetic and electrostatic analysers. These analysers filter out unwanted isotopes and molecules, allowing only the desired ions to reach the detector. By counting these ions, AMS can determine the isotopic composition of a sample, making it particularly useful for radiocarbon dating and tracing isotopic signatures in environmental and biological studies.

\noindent One of the key applications of AMS is in radiocarbon dating. As with all radiocarbon dating, as plants uptake C through photosynthesis, they take on the 14C activity of the atmosphere. Anything that derives from this Carbon will also have atmospheric 14C activity (including humans). If something stops actively exchanging Carbon (it dies, is buried, etc), that 14C begins to decay. Using AMS enables the use of very small sample sizes, compared to traditional carbon dating methods.

\section{Security applications}
Particle accelerators are increasingly being used for national security applications including explosive detection and nuclear threat reduction. A common usage is in cargo scanning at ports and airports, which often involves using high-energy X-ray beams produced by compact electron accelerators to inspect containers non-invasively. These systems can reveal hidden contraband or dangerous substances that may be smuggled within cargo shipments. 

\noindent In this application, accelerator-based sources of X-Rays (up to 6 MV) can be far more penetrating than Co-60 sources. Other technologies exist based on neutrons. Today, dual energy X-ray scanning is now frequently implemented, and these techniques are able to identify materials to high accuracy~\cite{Tang2016}.

\begin{figure}[h!]
    \centering
    \includegraphics[height=100pt]{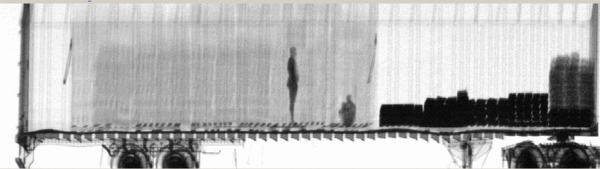}
    \caption{Cargo scanning image of a shipping container showing two stowaways hidden inside, Image: Public Domain.}
    \label{fig:10}
\end{figure}

\noindent In general, a container must be scanned in around 30 seconds for economic viability and throughput. This creates a tension between the flux or dose used and the resolution of the scan. One must also consider the possibility that a person may be in a truck or container: even if they are not supposed to be there, international radiation protection laws are strictly applied to limit the radiation dose such a person might receive. 

\section{Historical and cultural applications}

\noindent As previously discussed, IBA techniques in the MeV range can show art conservators and archaeologists the chemical composition of pigments used in paint, and backscattered radiation can give a detailed analysis of atoms present in a surface. This allows art historians to compare these measurements with paints known to be historically available to artists like Leonardo da Vinci. These techniques motivated the AGLAE facility, the only dedicated particle accelerator for art conservation, located in the basement of the famous Louvre museum in Paris.

\noindent Particle accelerators also play a crucial role in preserving cultural heritage artifacts. Non-destructive analysis techniques like X-ray fluorescence spectroscopy typically conducted at synchrotron light sources (see later) allow researchers to study the elemental composition of artworks without causing damage. This enables art historians and conservators to gain insights into artistic techniques, provenance verification, and identification of counterfeit objects.

\begin{figure}[h]
    \centering
    \includegraphics[height=150pt]{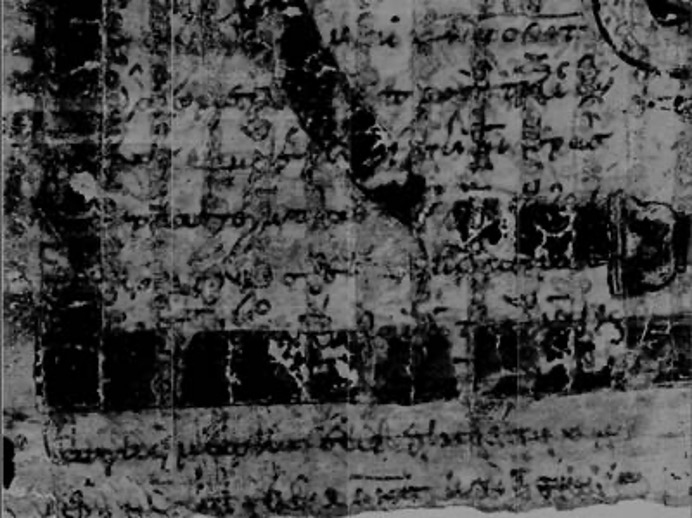}
    \caption{X-ray Fluorescence Image of an ancient work by Archimedes.}
    \label{fig:11}
\end{figure}

\noindent By way of example, X-ray fluorescence imaging carried out at SSRL revealed hidden text written by Archimedes by revealing the iron contained in the ink used by a 10th century scribe. The original work had been erased with lemon juice and written over during a period of shortage of parchment. The x-ray technique based image shows the lower left corner of the page~\cite{SLAC05}.

\section{Large Scale Scientific Facilities}
The design and operation of synchrotrons for large scale scientific facilities including synchrotron light sources and high power proton sources is covered elsewhere in the CERN Accelerator School course. Here we review details of techniques and common applications which are extremely wide ranging.

\subsection{Synchrotron light sources}
\begin{figure}[h]
    \centering
    \includegraphics[height=200pt]{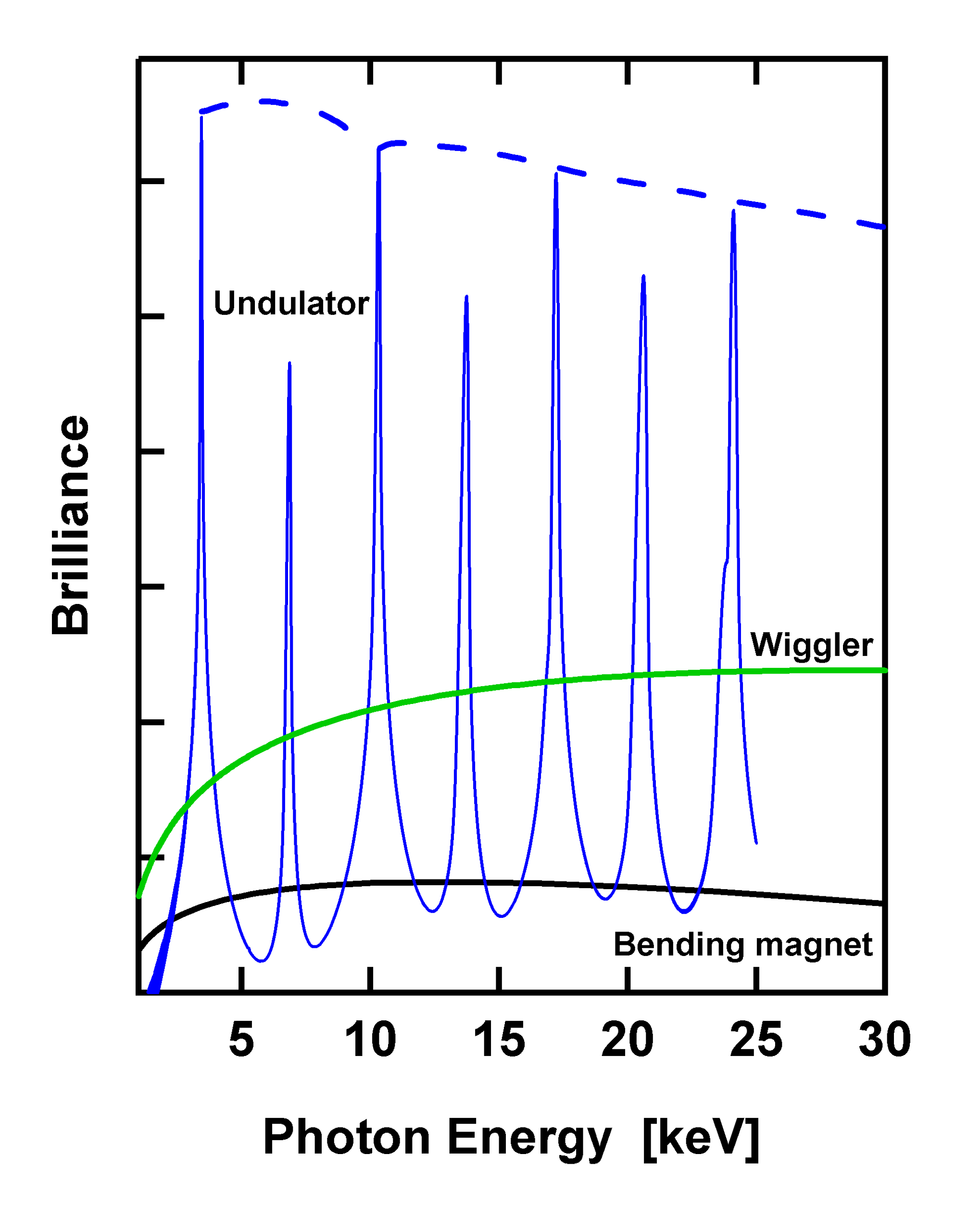}
    \caption{Brilliance of emitted photons for different insertion devices in a synchrotron light source, showing spectra for a bending magnets, wiggler and undulator.}
    \label{fig:12}
\end{figure}

Synchrotron radiation is emitted by charged particles when accelerated radially, or in other words, bent in a magnetic field. In a modern synchrotron light source radiation is typically produced using an insertion device: an undulator or wiggler. The choice of device depends on the intended application as the magnetic field and oscillatory period affects radiation produced. In an undulator, the radiation produced is coherent and tunable, but altering the undulator gap will vary the harmonic energies. While for a wiggler, the radiation produced is incoherent but high flux. Typical spectra are shown in Fig.\ref{fig:12}.

\noindent The light is directed down a series of beamlines and employed in a wide range of advanced techniques, including X-ray diffraction, X-ray absorption spectroscopy, and X-ray imaging, which are crucial for determining the atomic and molecular structures of materials. Techniques such as small-angle X-ray scattering (SAXS) and wide-angle X-ray scattering (WAXS) allow researchers to study nanostructures and macromolecular assemblies, while X-ray tomography provides three-dimensional imaging of internal structures without destructive sampling. Synchrotron light sources are instrumental in fields like materials science, biology, chemistry, and environmental science. They enable detailed studies of crystalline materials, biological macromolecules, catalysts, and complex fluids, aiding in the development of new materials, pharmaceuticals, and technologies. Additionally, synchrotron-based techniques support research in geology, archaeology, and cultural heritage, allowing non-destructive analysis of rare and valuable samples. The unparalleled brightness and precision of synchrotron light make these facilities indispensable for cutting-edge scientific research and technological innovation.

\noindent Some of the myriad application areas at synchrotron light sources include: condensed matter science, material science, engineering, chemistry, life sciences, structural biology, medicine, earth science, environmental science, cultural heritage and studies of instrumentation. 

\noindent To highlight just a few examples, consider the field of structural biology. Some examples of structural biology studies conducted at synchrotron light sources include protein crystallography, which recently allowed reconstruction of the main proteins of SARS-COV-2, while as far back as 1990 scientists determined the structure of a strain of foot and mouth virus using the Daresbury SRS. This has also been used for reconstruction of the 3D structure of a nucleosome (DNA packaging) with a resolution of 0.2 nm. The collection of precise information on the molecular structure of chromosomes and their components can improve the knowledge of how the genetic code of DNA is maintained and reproduced.  

\noindent There are excellent sources of further reading provided by synchrotron facilities including ESRF\cite{ESRF}.

\subsection{Spallation Neutron sources}
Spallation neutron sources are advanced facilities based usually on rapid cycling synchrotrons or high power linear proton accelerators. They are designed to generate neutrons through the spallation process, which involves striking a heavy metal target, such as tungsten, with a high-energy proton beam in the GeV energy range. When the high-energy protons collide with the nuclei of the target material, they cause the nuclei to become highly excited and emit a large number of neutrons. This process is efficient and can produce a high flux of neutrons, which are then slowed down, or moderated, to the desired energies using materials like water or liquid hydrogen.

\noindent The neutrons produced by spallation sources are invaluable in a wide range of scientific and industrial applications. In materials science, they are used to probe the atomic and magnetic structures of materials, providing insights into the properties and behaviours of metals, ceramics, polymers, and complex fluids. Neutron scattering techniques can reveal information about crystal structures, phase transitions, and magnetic ordering that is often inaccessible with other methods such as X-ray diffraction.

\noindent In chemistry and biology, spallation neutron sources facilitate the study of molecular dynamics and interactions. For example, they allow scientists to examine the structures of proteins, enzymes, and other biological macromolecules in detail, which is crucial for drug discovery and understanding biochemical processes. Neutrons are particularly suited for these studies because they interact with atomic nuclei rather than electron clouds, making them highly sensitive to light elements like hydrogen, which are abundant in biological systems.

\noindent Spallation neutron sources also play a significant role in fundamental physics research. They are used to investigate neutron properties, such as lifetime and magnetic moment. Moreover, these facilities contribute to advancements in nuclear physics by providing a means to study neutron-rich isotopes and nuclear reactions under controlled conditions.

\noindent Additionally, spallation neutron sources have practical applications in industry. They are used in non-destructive testing and analysis of materials and components, helping to ensure the safety and reliability of critical infrastructure, such as pipelines, aircraft, and nuclear reactors. The ability to analyze residual stresses and detect flaws within materials without causing damage is a significant advantage in quality control and failure analysis.

\noindent Their ability to produce intense neutron beams with precise control over energy and timing makes them indispensable for both fundamental research and practical applications. Accelerator-based neutron spallation sources are considered a complementary tool to synchrotron light sources. For some recent examples of scientific studies carried out using neutrons, a series of case studies are provided by the STFC ISIS neutron/muon source~\cite{ISIS-science}.

\section{Discussion and Challenges for Accelerator Applications}
As we have seen, particle accelerators have an astonishing number of uses that benefit society. However, it is essential to acknowledge the challenges and limitations associated with utilising particle accelerator technology outside of particle physics. One significant challenge lies in cost and complexity. Particle accelerators are complex machines that require substantial financial investment for construction, operation, and maintenance. Additionally, expertise in operating these instruments is crucial for obtaining accurate results and ensuring their safe usage. There also exist policy and regulation challenges, ethical considerations, and the potential misuse of technology in the wrong hands

\noindent One challenge recently highlighted with a new international collaboration is equitable access to radiotherapy technology worldwide.

\noindent Sustainability is another key consideration for societal applications of particle accelerators. Projects are growing in number worldwide to address sustainability issues of these technologies, as despite their net social good, it is more important than ever to consider environmental and other consequences of these technologies~\cite{Nature23}. New facilities must consider total electrical power consumption, make choices on the use of superconducting magnets, electromagnets or permanent magnets, energy recovery systems, and life-cycle analysis of any components that may become radioactive in the lifetime of a facility. Many accelerators use rare earth metals, for example Niobium in superconducting cavities and Neodymium or Samarium in permanent magnets: the recovery and recycling of these and other materials should be considered. 

\noindent Both radiotherapy linear accelerators and static voltage accelerators commonly used for ion beam analysis and radiocarbon dating use greenhouse gases for managing high voltages including SF6, and alternatives should be actively sought moving forward. Access to Helium for cryogenic systems is also of growing concern.

\noindent Finally, but of critical importance ethically: we must consider the social and cultural impacts of accelerators, and take care to implement best practice in the consultation and co-creation of new facilities in partnership with residents and local first nations populations.

\bibliographystyle{IEEEtran}
\bibliography{CASapplications-Suzie-latest}

\end{document}